\documentclass[final,2p,times,twocolumn,authoryear]{elsarticle}

\usepackage{amssymb}
\usepackage{amsmath}
\usepackage{bm}
\usepackage{xcolor}
\usepackage{lineno}  
\usepackage{hyperref}

\journal{Radiation Physics and Chemistry}

\begin{document}

\begin{frontmatter}

\title{Real-time portable muography with Hankuk Atmospheric-muon Wide Landscaping : HAWL}

\author[1]{J.~Seo}
\author[2]{N.~Carlin}
\author[2]{D.~F.~F.~S.~Cavalcante}
\author[1]{J.~S.~Chung}
\author[2]{L.~E.~França}
\author[1,b]{C.~Ha}
\ead{chha@cau.ac.kr}
\author[1]{J.~Kim}
\author[1]{J.~Y.~Kim}
\author[1]{H.~Kimku}
\author[1]{B.~C.~Koh}
\author[1]{Y.~J.~Lee}
\author[2]{B.~B.~Manzato}
\author[1]{S.~W.~Oh}
\author[2]{R.~L.~C.~Pitta}
\author[1]{S.~J.~Won}

\affiliation[1]{organization={Department of Physics, Chung-Ang University},
            city={Seoul},
            postcode={06974}, 
            country={Republic of Korea}}

\affiliation[2]{organization={Physics Institute, University of São Paulo},
            city={São Paulo},
            postcode={05508-090}, 
            country={Brazil}}

\begin{abstract}
  Cosmic ray muons prove valuable across various fields, from particle physics experiments to non-invasive tomography, thanks to their high flux and exceptional penetrating capability.
  Utilizing a scintillator detector, one can effectively study the topography of mountains situated above tunnels and underground spaces.
  The Hankuk Atmospheric-muon Wide Landscaping (HAWL) project successfully charts the mountainous region of eastern Korea by measuring cosmic ray muons with a detector in motion.
  The real-time muon flux measurement shows a tunnel length accuracy of 6.0\,\%, with a detectable overburden range spanning from 8~to 400~meter-water-equivalent depth.
  This is the first real-time portable muon tomography.
\end{abstract}

\begin{keyword}
Muon tomography\sep Plastic scintillator\sep Portable radiation detector



\end{keyword}

\end{frontmatter}

\section{Introduction}

A cosmic-ray muon is an elementary particle generated when a primary cosmic-ray particle collides with atmospheric nuclei~\cite{PDG-mu}. Cosmic-ray muons, in abundance, can traverse high-density materials non-destructively, and their unique energy loss in a material renders them valuable in various applications, from particle physics experiments to muon tomography.
In particle physics experiments, a high spatial and temporal resolution muon counter can measure muon flux, track final state particles in an accelerator beam~\cite{DUNE:2021tad}, and reveal yearly modulations with zenith angle dependence~\cite{Tilav:2019xmf}. Muon tomography has uncovered unknown spaces within pyramids~\cite{pyramid} and is employed to assess the condition of nuclear power plants~\cite{npp}. Furthermore, recent advancements in high-resolution muon imaging technology and portable detectors have broadened their applications, notably in volcanic activity detection~\cite{volcano}, archaeological site investigations~\cite{Avgitas:2022gls}, and underground safety measures~\cite{boringtomo,PhysRevResearch.2.023017}.
\begin{figure}[!htb]
  \begin{center}
    \includegraphics[width=0.48\textwidth]{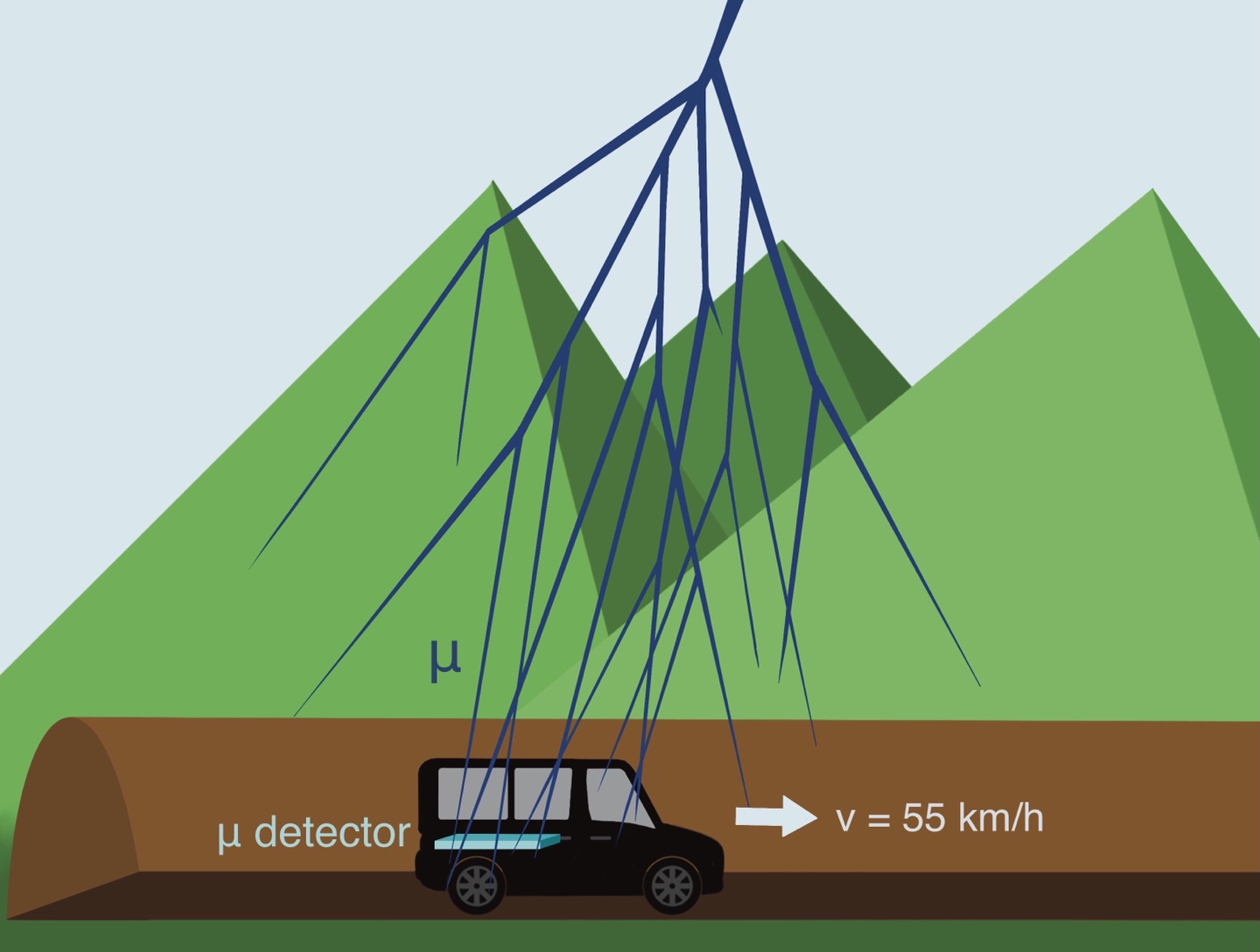}
  \end{center}
  \caption{Schematic diagram of the HAWL tomography measurement.
    The cosmic-ray muon flux is measured in real-time by the HAWL detector loaded in a moving vehicle.
    The change in the flux relative to the open area
    can represent the vertical overburden of the rock above the detector. 
  }
  \label{diagram}
\end{figure}

The primary goal of this research is to map the large-scale ($\sim$100~km) land forms and features above the underground spaces quickly
and non-invasively using a moving muon detector.
To reconstruct mountainous topography using a measured muon flux,
we launched the Hankuk Atmospheric-muon Wide Landscaping (HAWL) project.
HAWL precisely gauges changes in muon flux within tunnels situated above the Seoul-Yangyang highway
and Yangyang Underground Laboratory (Y2L)~\cite{Adhikari:2017esn} in South Korea
while moving at high speed and reporting results real time as schematically shown in Fig.~\ref{diagram}.
With data collected by the HAWL detector, we measured the muon flux as a function of elevation, and
the detector sensitivity in terms of depth and length resolutions for the tunnels.

HAWL contributes to the field of muography by providing technological and scientific innovations.
The muon detector is built by integrating multi-channel voltage supply, signal digitization,
and data acquisition in a single module under 68~Watt power consumption.
The 0.3~m$^2$ sized detector fits to a small car, operates in rugged conditions with high-speed dynamic movement. For the first time, a fast scan over hundred kilometers within a few hours has been achieved. Additionally, the real-time analysis and result reporting make the detector versatile and broadly applicable. Finally, the newly developed algorithms for data fitting enhance the detector's accuracy for measuring tunnel lengths and depths, demonstrating its full potential for larger multi-panel detectors.

\section{Experimental Method}
The main technical design goal of the HAWL experiment is to be compact enough to fit in a vehicle
and to be simply-integrated to consume low power.
We chose plastic scintillator (PS) as the main detection medium which is durable in rough conditions and relatively easy to handle~\cite{knoll}. 
To be efficient in power and space usage, small-sized silicon photomultipliers (SiPMs) coupled with optical fibers are used as the light sensor.
A newly developed, customized data acquisition system has been created in-house to streamline the overall data processing.

\subsection{Detector Construction}

The muon counter is constructed from rectangular PS material (Eljen Technology, EJ-200)~\cite{ejps} with dimensions of $\rm 490~mm \times 600~mm \times 30~mm$.
A total of 56 optical fibers coupled with 28 SiPMs are laid out as a grid on the top and bottom surfaces of the PS material.
  
The cylindrical fiber (Kuraray, double-cladding Y-11)~\cite{fiber} is a wavelength-shifting scintillator with 1~mm diameter that has an absorption peak at 430~nm and an emission peak at 476~nm with an attenuation length greater than 3.5~m. The SiPM (Hamamatsu, S13360-1375PE)~\cite{sipm} consists of 285 pixels and has a gain of $4.0 \times 10^6$ in a $\rm 1.3 \times 1.3~mm^2$ photosensitive area.
12 SiPMs are placed on the short sides and 16 on the long sides.
The arrangement included 56 optical fibers crossing each other at 19 mm intervals on both the top and bottom of PS with 4 ends of optical fibers joined together to couple one SiPM.
These fibers are firmly attached to the PS surfaces with transparent Scotch tape.
For optimal performance, the tip of each optical fiber was polished, and
a small amount of optical grease was applied at the junction between the SiPM and the optical fiber ends.

\begin{figure}[!htb]
  \begin{center}
      \includegraphics[width=0.48\textwidth]{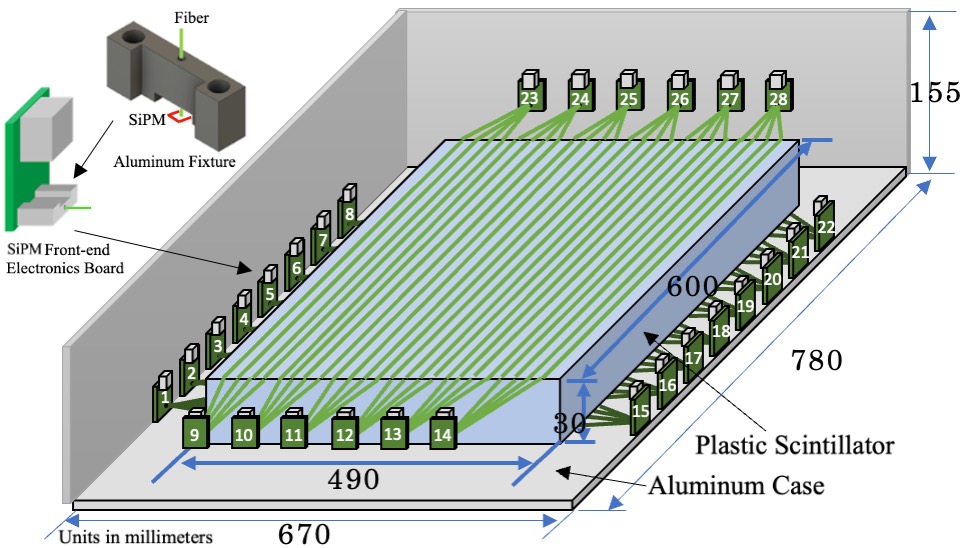}
  \end{center}
  \caption{The HAWL detector design. The optical fibers are laid out on top and bottom of a PS panel and
    their ends are coupled with 28 SiPM light sensors. 
    A SiPM is mounted on a front-end electronics board which sends signals to DAQ via LAN cables.
    A fixture aids the coupling between the SiPM sensitive area and the tip of the optical fibers.
    The entire setup is encased in an aluminum housing.
  }
  \label{blueprint}
\end{figure}

To better couple the tip of the optical fibers on a SiPM sensitive area,
a fixture as shown in Fig.~\ref{blueprint} is installed to the SiPM front-end board.
The optical fiber fixture has a hole with a diameter of 4.0~mm and a vertical depth of 8.0~mm
stabilizing the fiber bundle and providing better contact to the sensor.

\begin{figure*}[!htb]
  \begin{center}
      \includegraphics[width=1.0\textwidth]{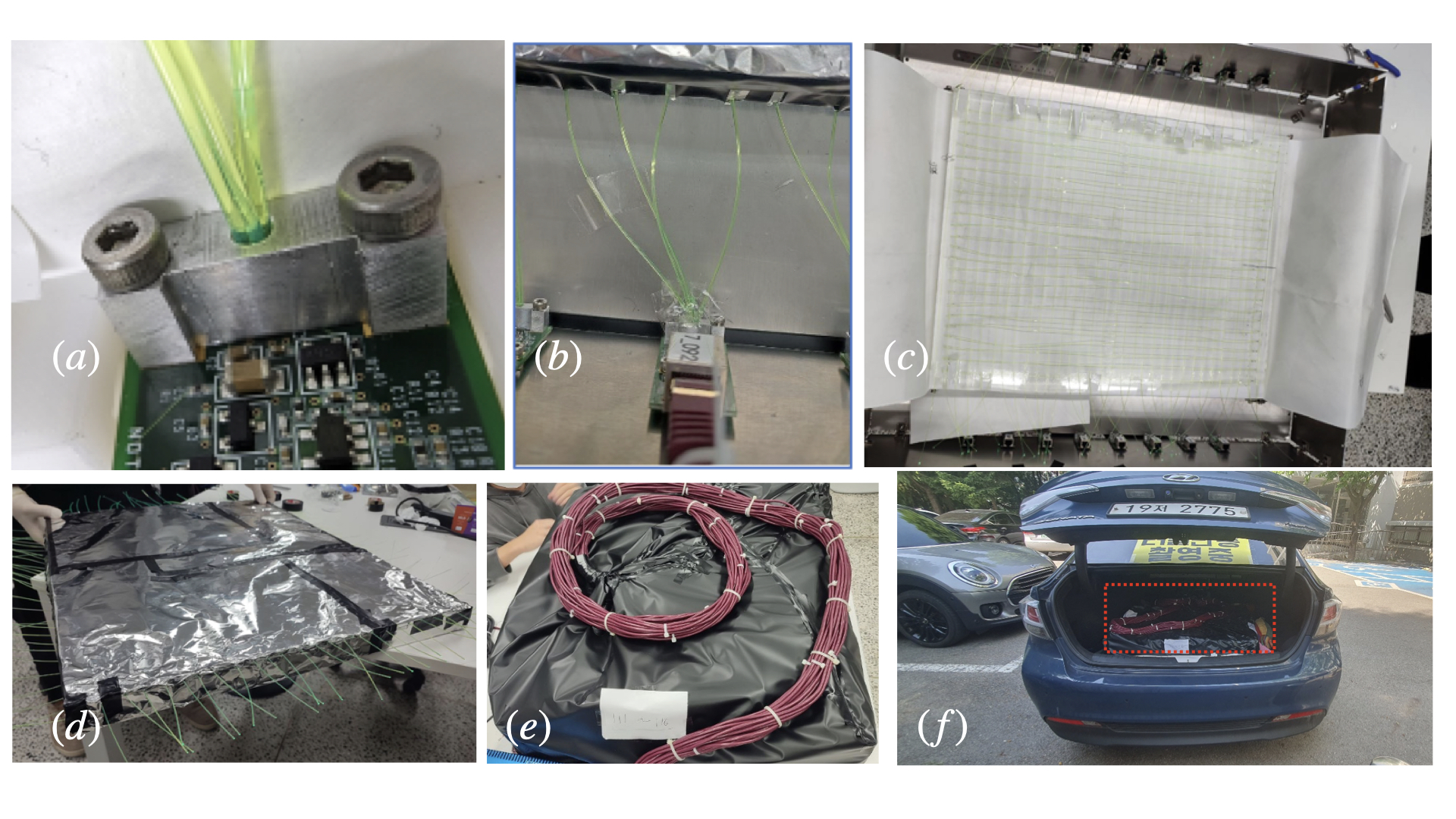}
  \end{center}
  \caption{Photos of the HAWL detector assembly.
    A front-end board connected with four fiber ends are shown in (a) and (b).
    The layout of the fibers with Tyvek sheets and additional aluminum foil covers are visible in (c) and (d),
    respectively.
    The completed HAWL detector is seen in (e) and
    it is set in place in the trunk of a sedan indicated by a red-dotted line (f) with the DAQ system in a back seat.
  }
  \label{hawl1}
\end{figure*}

To prevent leakage of scintillation photons and block external photons,
the panel with fibers is first wrapped in a Tyvek sheet which has more than 90\,\% reflectivity in the scintillation spectrum~\cite{tyvek}
and then is covered by a layer of 50~$\rm \mu m$-thick aluminium foil.
We placed the PS detector, assembled with SiPM front-end boards, inside a 3~mm-thick aluminium case wrapped in several layers of black sheets for added stability and radiation protection. A layout of the detector is shown in Fig.~\ref{blueprint} and its construction procedures are displayed in Fig.~\ref{hawl1}.

 \subsection{Data Acquisition}
\begin{figure}[!htb]
  \begin{center}
      \includegraphics[width=0.50\textwidth]{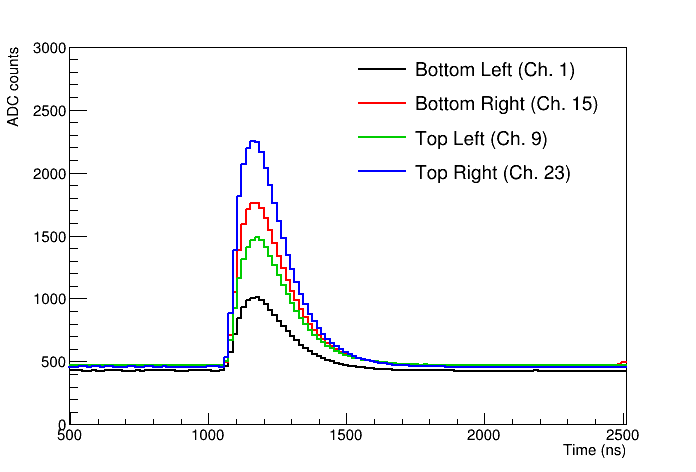}
  \end{center}
  \caption{SiPM raw waveform data. Four signal waveforms selected from each side of the detector are displayed.
    Baselines are typically set around 500~ADC and the amplitude can go up to 4096 ADC.
    The trigger time starts around 1000~ns and readout window goes to 4096~ns.
  }
  \label{wave}
\end{figure}

The data acquisition system consists of slow analog to digital converter (SADC), 
trigger--clock board (TCB) and a desktop computer~\cite{COSINE-100:2018rxe}. 
The 28 SiPM front-end boards are connected to a custom-made SADC module via a LAN connector.
The 8-pin LAN cable is to provide bias voltages and power with ground, and to read signals and temperatures for a single SiPM. 
The SADC module supplies required bias voltages to each channel and digitizes the analog signals.
The SADC uses 16~ns sampling and has two modes; full waveform mode and sum charge mode.
In the waveform mode, all channel waveforms per event are stored while in the sum charge mode,
reduced information consisting of only integrated ADC values and average time values for each channel are stored.
Example waveforms for the same signal event are shown in Fig.~\ref{wave}.

The time synchronization and trigger decision are performed by the TCB
which is connected to the SADC module via a USB cable.
A channel hit is defined when a SiPM signal exceeds the integral charge threshold of 200 ADCs in a 100~ns sliding window on that channel. A trigger is formed if 8 or more hits out of 28 active channels are recorded within a 1000~ns time window, and an event is constructed by padding the trigger time around a 4~$\mu$s readout window, aligning the trigger hit times near 1.2~$\mu$s. The minimum threshold of 8 was determined based on a trigger rate tolerance of 2~kHz and a minimum energy threshold of approximately 1 MeV. This is compared with the expected muon rate of 50~Hz and the average deposited muon energy of 6~MeV, providing sufficient margins.

The collected raw data are in a binary format which is immediately converted into a ROOT format~\cite{root}. 
The detector tests are done in a waveform mode
where the trigger settings and the individual channel signal analysis are determined in advance.
Then, those conditions are applied to the real-time physics runs using a sum charge mode
to minimize the power and disc space loads.

\begin{figure}[!htb]
  \begin{center}
      \includegraphics[width=0.5\textwidth]{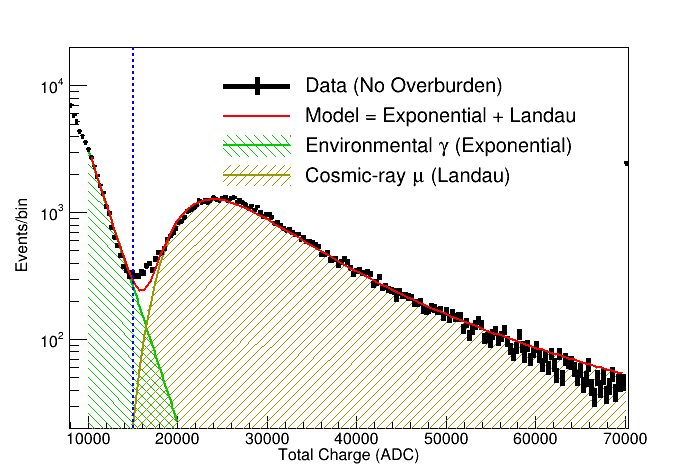}
  \end{center}
  \caption{Muon event selection.
    The rate measurements in an open space are shown as a function of 28-SiPM sum charge.
    Data are fit to a model defined by an exponential component for environmental radiation (green fills) plus
    a Landau component for cosmic-ray muons (yellow fills).
    The vertical dotted line at 15000~ADC is the cut to select muon events.
  }
  \label{events}
\end{figure}

The completed detector has been tested in a lab to set a muon selection cut
and check the stability before beginning the physics run collections.
Figure~\ref{events} shows event distributions in terms of their measured charge in open space.
Muons are well separated from the environmental gamma background.
A Landau function plus an exponential function was used to model the data for muon and gamma components, respectively.
Based on the model fit, a muon selection cut was set at 99\,\% estimated signal efficiency, allowing 2\,\% gamma contamination.
Subsequently, all equipment was loaded onto the vehicle with an 860~Watt-hour battery.
The real-time monitoring was established, allowing results to be broadcasted via WIFI.
This includes recordings of the speed of the car, the GPS locations of tunnels, and tunnel images.

Muon flux measurements have been performed twice with the same portable HAWL detector on November 25, 2022 (TRIP-1) and June 12, 2024 (TRIP-2).
The trips began from the Gapyeong service area which is located approximately 60~km east of Seoul in the Seoul-Yangyang national highway and traveled to the easterly direction.
The muon event rate as a function of time has been displayed in the car.
At the same time, the same event rate is broadcasted in real-time by Internet so that
it can be simultaneously monitored remotely.

\begin{figure}[!htb]
  \begin{center}
      \includegraphics[width=0.45\textwidth]{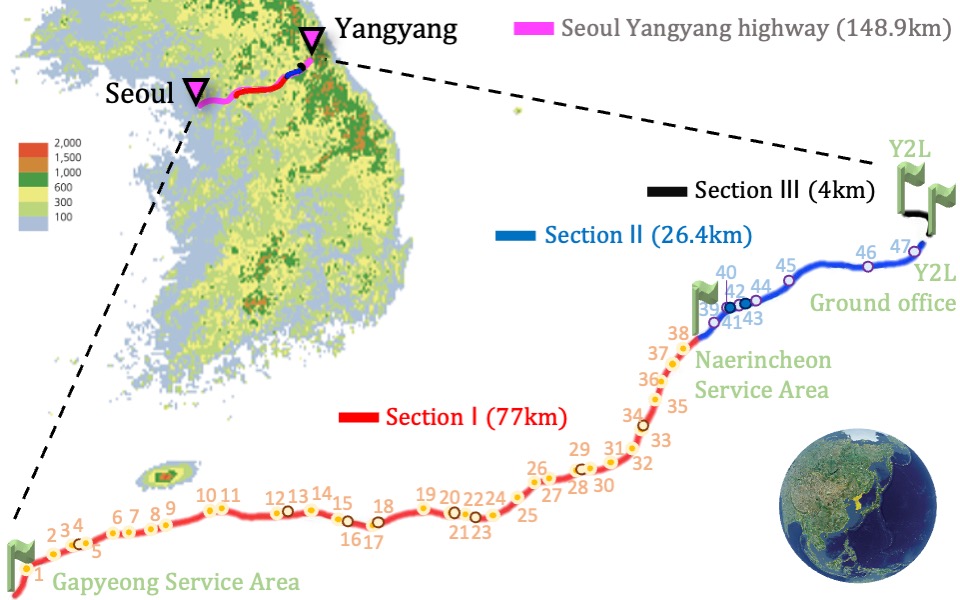}
  \end{center}
  \caption{Measurement campaign sections and tunnels.
    The trip was divided into three sections (I,II, and III).
    Section I (red line) is from Gapyeong service area to Naerincheon service area
    and Section II (blue line) is to Y2L ground office.
    Section III (black line) is on a local road from Y2L ground office to Y2L underground lab.
    Tunnels are numbered in sequence. The specification of each tunnel can be found in Table~\ref{tunnelspec}.
  }
  \label{map}
\end{figure}

These campaigns are divided into three sections based on their geographical uniqueness. 
Section~I consists of many tunnels of varying lengths and types
while Section~II
includes the longest tunnel.
We enter Yangyang underground laboratory (Y2L) which is the 700~m deep facility that can be accessed by a car in Section~III.
These sections covered about 100~km out of the total length of the 150~km highway as shown in Fig.~\ref{map}.
We maintained the speed at 55~km/h for TRIP-1 and 51~km/h for TRIP-2 (the highway lanes have low speed limit of 50~km/h).

\section{Analysis and Results}

Physics data of 10822 and 12475 seconds have been collected for TRIP-1 and TRIP-2, respectively. 
The run-times for each section and average rates are summarized in Table~\ref{campaign}.
The moment the car enters a tunnel, the HAWL detector is able to show the muon rate decrease visually as displayed in Fig.~\ref{hawl3} and
in a video of the Supplementary material. 

\begin{table}[ht]
  \begin{center}
    \caption{
      Information for the two HAWL travels.
    }
  \label{campaign}
  \begin{tabular}{lccccccc}
    \hline                                                                                
    Trip(month)   & Run-times(I/II/III) & Rates(raw)\\
    \hline                                                                 
      1(2022-Nov.) &	5259~s/2475~s/3088~s  & 1901.5~Hz\\
      2(2024-Jun.)   &	7213~s/2883~s/2379~s  & 224.3~Hz  \\
      \hline  
  \end{tabular}
  \end{center}
\end{table}

Figure~\ref{hawl3} shows the muon event rates as a function of time
for Section I and II for TRIP-1 and TRIP-2. 
The HAWL detector was able to delineate all tunnels, overpasses, and features above the moving car.
The average muon rate was measured at $53.0 \pm 0.1$~Hz throughout the campaign.
It was reduced when passing a tunnel depending on its length and depth.
Section-II includes the 11~km-long tunnel (Inje-Yangyang tunnel) where the maximum vertical overburden is 440~m.
Data show a clear suppression in rates inside the tunnels where we were able to measure the tunnel width and a maximum depth of overburden.
The lab test at 18~m above sea level shows a muon flux of 154.9~events\,$\rm /m^{2}/s$.
\begin{figure*}[!htb]
  \begin{center}
      \includegraphics[width=1.05\textwidth]{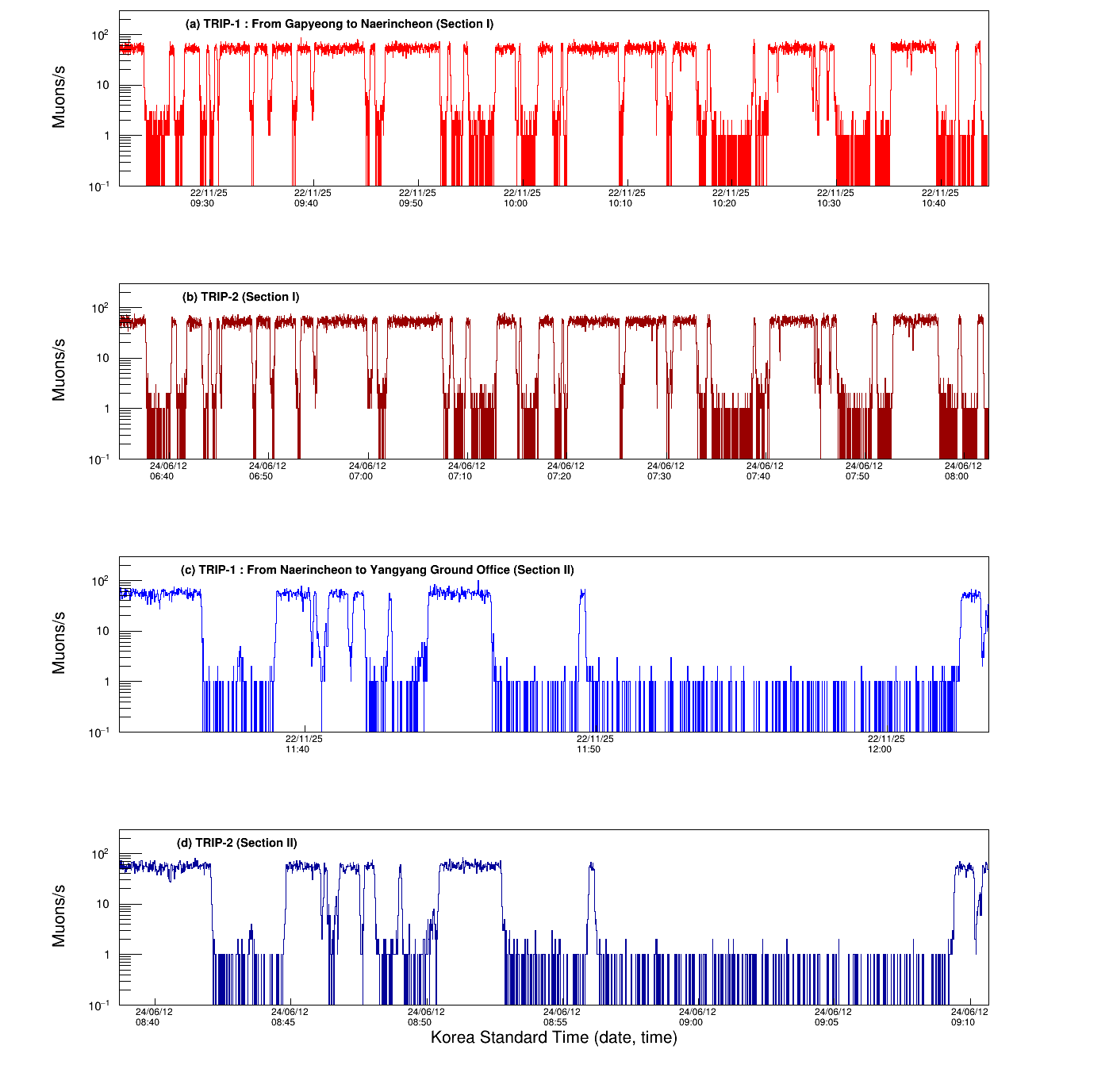}
  \end{center}
  \caption{Muon flux measurement campaigns.
    Muon count rates per second are displayed in Section I(a,b) and II(c,d) for TRIP-1 (a,c). and TRIP-2 (b,d).
    The rate drops are shown when passing tunnels.
    Note that the low count rates at deep tunnels in plot (c) and plot (d)
    appear to be mostly gamma background events.
  }
  \label{hawl3}
\end{figure*}

\begin{figure}[!htb]
  \begin{center}
      \includegraphics[width=0.5\textwidth]{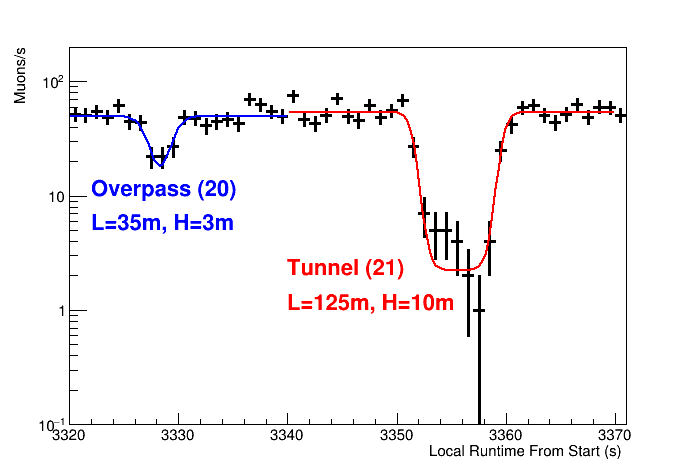}
  \end{center}
  \caption{Example short tunnel measurements (TRIP-1).
    Two short tunnels are fitted separately.
    The blue fit line is for the 35~m-long animal overpass with height of 3~m while
    the red fit line is for the 125~m-long tunnel with maximum height of 10~m.
  }
  \label{fit}
\end{figure}

We model the muon flux reduction based on the extended Sigmoid function $f(t)$
which is defined by
\begin{equation} \label{eq}
  f(t) = \frac{A}{1+e^{b(t-m)}} + \frac{A}{1+e^{-b(t-m-w)}} + d, 
\end{equation}
where $A$ describes open area muon rate, $b$ models the slope of the mountain,
$m$ and $w$ represent entrance and width of the tunnel, respectively, and $d$
is the depth parameter. The first term of Eq.~\ref{eq} is for the entrance side of the tunnel while
the second term is mirror reflection for the exiting side. 
The data segments that show rate reductions were fitted with Eq.~\ref{eq} using the $\chi^2$ minimization
to get the best-fit parameters and their uncertainties.
An example fit is shown in Fig.~\ref{fit}.

The final results presented are based on the statistical average of the independent analyses from two trips. We performed separate fittings for each trip’s data first
and then combined the best-fit values into a final result.

\begin{figure}[!htb]
  \begin{center}
      \includegraphics[width=0.48\textwidth]{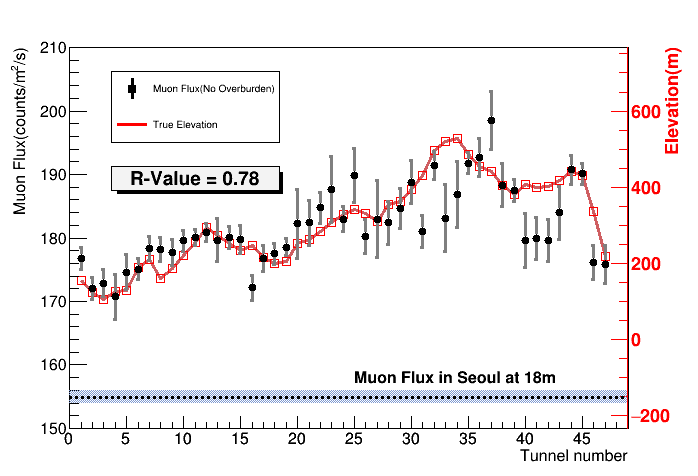} 
  \end{center}
  \caption{Open space muon flux as a function of the tunnel location.
    The muon rate is closely following the elevation changes of the Seoul-Yangyang highway.
    The correlation coefficient is measured to be $R=0.78$.
    Please note that the dotted line is the reference muon flux
    measured at 18~m above sea level with the same detector.
    Data points are average values from the two trips.
  }
  \label{mu_change}
\end{figure}

The fitting results show correct identification of all tunnels
when we compared our measurements with the original civil engineering data
and satellite images.
The short overpasses are identified with limited data points which
are constrained by the speed of the vehicle and the size of the detector. 
A model fit based on the Eq.~\ref{eq} reveals continuous measurements of the muon flux in open spaces
in between the tunnels.
Because the highway has been built on an increasing slope (maximum elevation difference is about 400~m) towards the end of Section I and II
and a decreasing slope at the end of Section II,
the muon rate appears to follow the similar characteristics of the elevation.
Figure~\ref{mu_change} shows the muon flux in an open space as a function of elevation.
A constant muon flux model is rejected at more than 5 standard deviations. 
The correlation coefficient $R=0.78$ between the muon flux and engineering elevation data
has been obtained supporting the flux increase as a function of the altitude.

\begin{figure}[!htb]
  \begin{center}
      \includegraphics[width=0.48\textwidth]{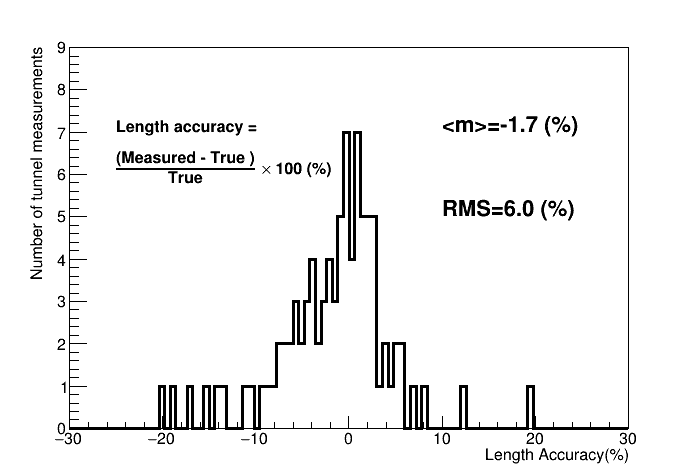} 
  \end{center}
  \caption{Length accuracy distribution.
    The relative difference between measured and actual lengths of tunnels is drawn without averaging of two trips.
    The accuracy is measured to be 6.0\,\% while the bias is -1.7\,\%.
    The length accuracy is driven by the outliers which are mostly short tunnels.
  }
  \label{length}
\end{figure}

Assuming the constant speed of the car (55~km/h and 51~km/h), we estimated
the tunnel entrance location and the width as shown in Fig.~\ref{length}.
The width accuracy was measured as 6.0\,\% by the spread of the relative difference
between the measured and true length. 

\begin{figure}[!htb]
  \begin{center}
      \includegraphics[width=0.48\textwidth]{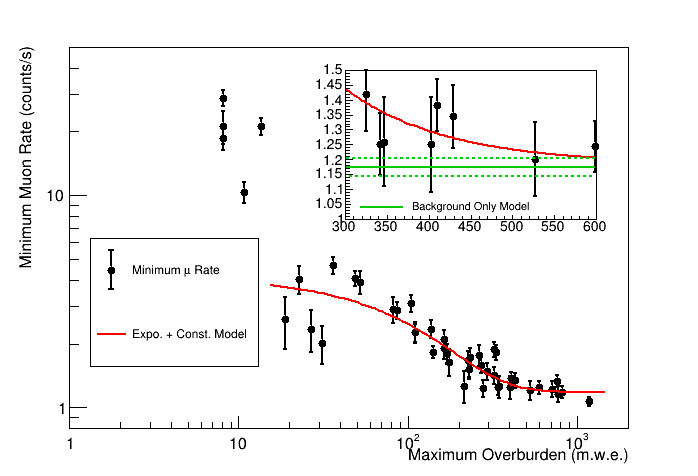} 
  \end{center}
  \caption{Measured minimum muon rate as a function of true maximum overburden in the meter-water-equivalent (m.w.e.) unit.
    The muon rate drops exponentially until it is saturated by the gamma contamination.
    Above 400~m.w.e. deep tunnel, the events are dominated by the background gammas.
    Note that the cluster of five points at around 10~m.w.e. corresponds to short animal overpasses which
    do not fit well to the exponential function (red line)
    due to their flat geometrical shape different from that of mountains.
    Inset plot shows a zoomed in view between 300 and 600~m.w.e. in linear scale where
    a background only model (solid green line) is drawn with 68\% error lines (dotted green).
    Data points are average values from the two trips.
  }
  \label{depth}
\end{figure}

The muon rates inside the tunnels have been measured as a function of their true maximum depths.
Assuming the metamorphic rock contents ($\rho = 2.7\,g/cm^3$) in northeastern Korean provinces,
the depth is converted into the meter-water-equivalent depth. 
Figure~\ref{depth} shows the expected exponential reduction behavior~\cite{PDG-mu}.
A function of an exponential part for muon rates and a constant part for background
is used to fit the data. 
The minimum overburden that data recorded is 3 meters (8.1~m.w.e.) while
the maximum overburden is determined at 148 meters (400~m.w.e.) by comparing with background only model.
At 400~m.w.e., the background-only hypothesis is rejected at 3.9 standard deviation
when compared to the best fit of the combined model.
The tunnel specifications and measured values are tabulated in Table~\ref{tunnelspec}.

\begin{figure}[!htb]
  \begin{center}
      \includegraphics[width=0.48\textwidth]{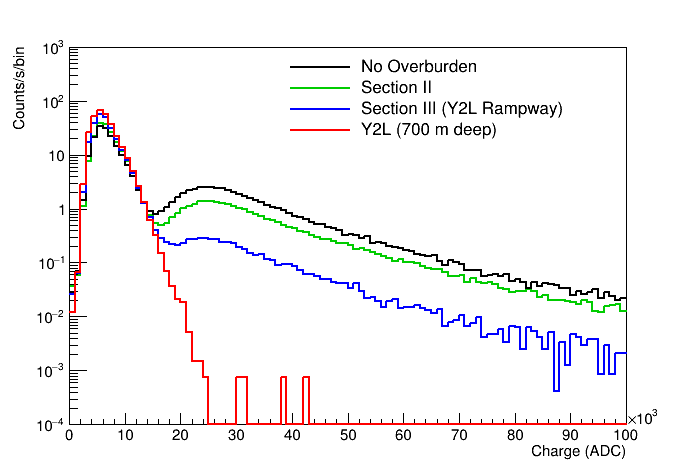} 
  \end{center}
  \caption{
    Charge spectra at different locations measured by HAWL TRIP-2.
    Section II (green) and Section III (blue) exhibit a clear overall reduction in muon rates compared to the open space measurement (black). The measurement at 700 meters deep in Y2L (red) indicates the largest suppression of cosmic-ray muons.
  }
  \label{secIII}
\end{figure}

In Section III measurements, we entered Y2L through a 2~km rampway at a speed of less than 5~km/h.
The muon rates are quickly reduced compared to the unchanged gamma background.
During an additional 1343 seconds of measurement at a depth of 700~m within Y2L,
we identified 4 muon candidate events, as shown in Fig.~\ref{secIII}.
This yields a rate of $(10.1\pm5.0)\times10^{-7}$/cm$^2$/s, comparable to past measurements of $(3.80\pm0.11)\times10^{-7}$/cm$^2$/s and $(4.4\pm0.3)\times10^{-7}$/cm$^2$/s in nearby laboratories for dark matter experiments within Y2L~\cite{COSINE-100:2020jml,KIMSmuonflux}.

\section{Discussion and Conclusion}

The primary challenge lies in acquiring muon data more efficiently, especially considering the speed limits on highways
and therefore, a bigger size panel is preferable.
Secondly, the separation power between the environmental gammas and muons
determines the depth resolution.
A detailed analysis after the return of the trip shows most of those
events recorded under deeper tunnels (i.e. depths larger than 148 meters) are
dominated by gamma background events.
Due to the relatively fast speed of the car and the small area detector,
it is not efficient to collect a large amount of muon data in a deeper tunnel.
On the other hand, we were able to map the short-to-medium tunnels with better accuracy thanks to high count rates.
The five shortest structures detected were animal overpasses that have less than 5~m overburden.
Their flat shapes different from regular mountains give less reduction of muons.
Additionally, data show that the entrance or exit of the tunnels can be measured with a high precision which helps
determine the length of the tunnel. The steepness of mountains near the entrance or exit of the tunnels
could be estimated utilizing the slope parameter in the fit.

The HAWL portable muon detector, utilizing a plastic scintillator, was successfully developed to create a mountainous topography, demonstrating muon flux variations based on rock thickness.
We have developed simplified data acquisition system and real-time reporting within a single module under 68~Watt power consumption.

A fast scan over hundred kilometers within a few hours has been achieved for the first time.
Two trips over the same range show consistent results.
The completed tests validated the detector's functionality, evident in the expected decrease
in muon flux when passing through a tunnel beneath a mountain.
Analyzing muon flux data along the Seoul-Yangyang highway revealed a close alignment with anticipated tunnel dimensions and tomography above it, achieving a length accuracy of 6.0\,\% for tunnels at depths ranging from 8 to 400~m.w.e.
An upgraded version of HAWL is in development, featuring two thicker panel detectors with anti-coincidence and directional sensitivity.
Given its versatility in shallow-depth applications, the HAWL detector is planned to be used in subway systems and underwater tunnels to enhance safety measures.

\section{Acknowledgments}
This work was supported by National Research Foundation of Korea~(NRF) grant funded by the Korean government~(MSIT)
(NRF-2021R1A2C1013761) and under the framework of international cooperation program managed by the National Research Foundation of Korea~(NRF-2021K2A9A1A0609413312), Republic of Korea;
Grant No. 2021/06743-1, 2022/12002-7 and 2022/13293-5 FAPESP, CAPES Finance Code 001, CNPq 304658/2023-5 and 303122/2020-0, Brazil. 

\bibliographystyle{elsarticle-harv}

\begin{thebibliography}{19}
\expandafter\ifx\csname natexlab\endcsname\relax\def\natexlab#1{#1}\fi
\providecommand{\url}[1]{\texttt{#1}}
\providecommand{\href}[2]{#2}
\providecommand{\path}[1]{#1}
\providecommand{\DOIprefix}{doi:}
\providecommand{\ArXivprefix}{arXiv:}
\providecommand{\URLprefix}{URL: }
\providecommand{\Pubmedprefix}{pmid:}
\providecommand{\doi}[1]{\href{http://dx.doi.org/#1}{\path{#1}}}
\providecommand{\Pubmed}[1]{\href{pmid:#1}{\path{#1}}}
\providecommand{\bibinfo}[2]{#2}
\ifx\xfnm\relax \def\xfnm[#1]{\unskip,\space#1}\fi
\bibitem[{Adhikari et~al.(2018a)}]{Adhikari:2017esn}
\bibinfo{author}{Adhikari, G.}, et~al., \bibinfo{year}{2018}a.
\newblock \bibinfo{title}{{Initial Performance of the COSINE-100 Experiment}}.
\newblock \bibinfo{journal}{Eur. Phys. J. C} \bibinfo{volume}{78},
  \bibinfo{pages}{107}.
\newblock \DOIprefix\doi{10.1140/epjc/s10052-018-5590-x},
  \href{http://arxiv.org/abs/1710.05299}{{\tt arXiv:1710.05299}}.
\bibitem[{Adhikari et~al.(2018b)}]{COSINE-100:2018rxe}
\bibinfo{author}{Adhikari, G.}, et~al. (\bibinfo{collaboration}{COSINE-100}),
  \bibinfo{year}{2018}b.
\newblock \bibinfo{title}{{The COSINE-100 Data Acquisition System}}.
\newblock \bibinfo{journal}{JINST} \bibinfo{volume}{13},
  \bibinfo{pages}{P09006}.
\newblock \DOIprefix\doi{10.1088/1748-0221/13/09/P09006},
  \href{http://arxiv.org/abs/1806.09788}{{\tt arXiv:1806.09788}}.
\bibitem[{Avgitas et~al.(2022)Avgitas, Elles, Goy, Karyotakis and
  Marteau}]{Avgitas:2022gls}
\bibinfo{author}{Avgitas, T.}, \bibinfo{author}{Elles, S.},
  \bibinfo{author}{Goy, C.}, \bibinfo{author}{Karyotakis, Y.},
  \bibinfo{author}{Marteau, J.}, \bibinfo{year}{2022}.
\newblock \bibinfo{title}{{Muography applied to archaelogy}}, in:
  \bibinfo{booktitle}{{27e \'edition de la R\'eunion des Sciences de la
  Terre}}.
\newblock \href{http://arxiv.org/abs/2203.00946}{{\tt arXiv:2203.00946}}.
\bibitem[{Brun and Rademakers(1997)}]{root}
\bibinfo{author}{Brun, R.}, \bibinfo{author}{Rademakers, F.},
  \bibinfo{year}{1997}.
\newblock \bibinfo{title}{{ROOT - An Object Oriented Data Analysis Framework}}.
\newblock \bibinfo{journal}{Nuclear Instruments and Methods in Physics Research
  Section A: Accelerators, Spectrometers, Detectors and Associated Equipment}
  \bibinfo{volume}{389}, \bibinfo{pages}{81}.
\bibitem[{Chevalier et~al.(2019)}]{boringtomo}
\bibinfo{author}{Chevalier, A.}, et~al., \bibinfo{year}{2019}.
\newblock \bibinfo{title}{Using mobile muography on board a tunnel boring
  machine to detect man-made structures}.
\newblock \bibinfo{journal}{AGU Fall Meeting Abstracts} \bibinfo{volume}{Vol.
  2019}.
\bibitem[{Eljen(2023)}]{ejps}
\bibinfo{author}{Eljen}, \bibinfo{year}{2023}.
\newblock \bibinfo{title}{Eljen Technology, GENERAL PURPOSE EJ-200, EJ-204,
  EJ-208, EJ-212}.
\newblock \URLprefix
  \url{https://eljentechnology.com/products/plastic-scintillators/ej-200-ej-204-ej-208-ej-212}.
\bibitem[{Hewes et~al.(2021)}]{DUNE:2021tad}
\bibinfo{author}{Hewes, V.}, et~al. (\bibinfo{collaboration}{DUNE}),
  \bibinfo{year}{2021}.
\newblock \bibinfo{title}{{Deep Underground Neutrino Experiment (DUNE) Near
  Detector Conceptual Design Report}}.
\newblock \bibinfo{journal}{Instruments} \bibinfo{volume}{5},
  \bibinfo{pages}{31}.
\newblock \DOIprefix\doi{10.3390/instruments5040031},
  \href{http://arxiv.org/abs/2103.13910}{{\tt arXiv:2103.13910}}.
\bibitem[{Janecek(2012)}]{tyvek}
\bibinfo{author}{Janecek, M.}, \bibinfo{year}{2012}.
\newblock \bibinfo{title}{Reflectivity spectra for commonly used reflectors}.
\newblock \bibinfo{journal}{IEEE Transactions on Nuclear Science}
  \bibinfo{volume}{59}, \bibinfo{pages}{490--497}.
\newblock \DOIprefix\doi{10.1109/TNS.2012.2183385}.
\bibitem[{Knoll(2010)}]{knoll}
\bibinfo{author}{Knoll, G.F.}, \bibinfo{year}{2010}.
\newblock \bibinfo{title}{Radiation detection and measurement}.
\newblock \bibinfo{publisher}{John Wiley and Sons, New York}.
\bibitem[{Kuraray(2024)}]{fiber}
\bibinfo{author}{Kuraray}, \bibinfo{year}{2024}.
\newblock \bibinfo{title}{Plastic scintilating fibers (PSF)}.
\newblock \URLprefix \url{https://www.kuraray.com/products/psf}.
\bibitem[{MPPC(2024)}]{sipm}
\bibinfo{author}{MPPC}, \bibinfo{year}{2024}.
\newblock \bibinfo{title}{MPPC for precision measurement, S13360-1375PE}.
\newblock \URLprefix
  \url{https://www.hamamatsu.com/eu/en/product/optical-sensors/mppc}.
\bibitem[{Prihtiadi et~al.(2021)}]{COSINE-100:2020jml}
\bibinfo{author}{Prihtiadi, H.}, et~al. (\bibinfo{collaboration}{COSINE-100}),
  \bibinfo{year}{2021}.
\newblock \bibinfo{title}{{Measurement of the cosmic muon annual and diurnal
  flux variation with the COSINE-100 detector}}.
\newblock \bibinfo{journal}{JCAP} \bibinfo{volume}{02}, \bibinfo{pages}{013}.
\newblock \href{http://arxiv.org/abs/2005.13672}{{\tt arXiv:2005.13672}}.
\bibitem[{Procureur et~al.(2023a)}]{npp}
\bibinfo{author}{Procureur, S.}, et~al., \bibinfo{year}{2023}a.
\newblock \bibinfo{title}{{3D imaging of a nuclear reactor using muography
  measurements}}.
\newblock \bibinfo{journal}{Science Advances} \bibinfo{volume}{9},
  \bibinfo{pages}{eabq8431}.
\bibitem[{Procureur et~al.(2023b)}]{pyramid}
\bibinfo{author}{Procureur, S.}, et~al., \bibinfo{year}{2023}b.
\newblock \bibinfo{title}{{Precise characterization of a corridor-shaped
  structure in Khufu’s Pyramid by observation of cosmic-ray muons}}.
\newblock \bibinfo{journal}{Nature Communications} \bibinfo{volume}{14},
  \bibinfo{pages}{1144}.
\bibitem[{Thompson et~al.(2020)Thompson, Stowell, Fargher, Steer, Loughney,
  O'Sullivan, Gluyas, Blaney and Pidcock}]{PhysRevResearch.2.023017}
\bibinfo{author}{Thompson, L.F.}, \bibinfo{author}{Stowell, J.P.},
  \bibinfo{author}{Fargher, S.J.}, \bibinfo{author}{Steer, C.A.},
  \bibinfo{author}{Loughney, K.L.}, \bibinfo{author}{O'Sullivan, E.M.},
  \bibinfo{author}{Gluyas, J.G.}, \bibinfo{author}{Blaney, S.W.},
  \bibinfo{author}{Pidcock, R.J.}, \bibinfo{year}{2020}.
\newblock \bibinfo{title}{Muon tomography for railway tunnel imaging}.
\newblock \bibinfo{journal}{Phys. Rev. Res.} \bibinfo{volume}{2},
  \bibinfo{pages}{023017}.
\newblock \URLprefix
  \url{https://link.aps.org/doi/10.1103/PhysRevResearch.2.023017},
  \DOIprefix\doi{10.1103/PhysRevResearch.2.023017}.
\bibitem[{Tilav et~al.(2020)Tilav, Gaisser, Soldin and Desiati}]{Tilav:2019xmf}
\bibinfo{author}{Tilav, S.}, \bibinfo{author}{Gaisser, T.K.},
  \bibinfo{author}{Soldin, D.}, \bibinfo{author}{Desiati, P.}
  (\bibinfo{collaboration}{IceCube}), \bibinfo{year}{2020}.
\newblock \bibinfo{title}{{Seasonal variation of atmospheric muons in
  IceCube}}.
\newblock \bibinfo{journal}{PoS} \bibinfo{volume}{ICRC2019},
  \bibinfo{pages}{894}.
\newblock \DOIprefix\doi{10.22323/1.358.0894},
  \href{http://arxiv.org/abs/1909.01406}{{\tt arXiv:1909.01406}}.
\bibitem[{Tioukov et~al.(2019)}]{volcano}
\bibinfo{author}{Tioukov, V.}, et~al., \bibinfo{year}{2019}.
\newblock \bibinfo{title}{{First muography of Stromboli volcano}}.
\newblock \bibinfo{journal}{Scientific Reports} \bibinfo{volume}{9},
  \bibinfo{pages}{6695}.
\bibitem[{Workman et~al.(2022)}]{PDG-mu}
\bibinfo{author}{Workman, R.}, et~al. (\bibinfo{collaboration}{Particle Data
  Group}), \bibinfo{year}{2022}.
\newblock \bibinfo{title}{{The Review of Particle Physics}}.
\newblock \bibinfo{journal}{Prog. Theor. Exp. Phys.} \bibinfo{volume}{2022},
  \bibinfo{pages}{083C01}.
\bibitem[{Zhu et~al.(2005)}]{KIMSmuonflux}
\bibinfo{author}{Zhu, J.}, et~al. (\bibinfo{collaboration}{KIMS}),
  \bibinfo{year}{2005}.
\newblock \bibinfo{title}{{Study on the Muon Background in the Underground
  Laboratory of KIMS}}.
\newblock \bibinfo{journal}{High Energy Physics and Nuclear Physics}
  \bibinfo{volume}{29}, \bibinfo{pages}{721}.

\end{thebibliography}

\appendix

\begin{table*}[ht]
  \begin{center}
    \caption{
      Tunnel specifications and HAWL measurements.
      Lengths (L) and maximum depths (H) are obtained from engineering blueprints.
      The elevations are an average value between the entrance and exit.
      Measured L is equal to the fit width parameter $w$ multiplied by the car speed.
      Depth H is the fit depth parameter $d$ and the muon flux is for open space rates before entering each tunnel.
      All measured values are average values from the two trips.
    }
  \label{tunnelspec}
  \begin{tabular}{lccccccc}
    \hline                                                                                
    No. &       Name       & L (m) & Max. H (m) & Elev. (m) & Meas. L(m) & Depth H(/s)  &  $\mu$ Flux($\rm /m^2/s$)\\
    \hline                                                                                
      1    &   Misa            &	2171	 &  284   &       155     & 2213.7 & 1.33   &176.7 \\
      2    &   Magok           &	919 	 &  85    &       124     & 903.0  & 1.52   &172.0 \\
      3    &   Balsan 1        &	633 	 &  84    &       106     & 619.0  & 1.67   &172.9 \\
      4    &   Balsan 2        &	480 	 &  97.5  &       126     & 455.4  & 1.76   &170.7 \\
      5    &   Balsan 3        &	254 	 &  30.5  &       129     & 213.2  & 2.89   &174.6 \\
      6    &   Balsan 4        &	433 	 &  60.5  &       189     & 421.4  & 1.91   &175.0 \\
      7    &   Chugok          &	433 	 &  32    &       211     & 426.7  & 2.87   &178.3 \\
      8    &   Haengchon       &	463 	 &  63.5  &       160     & 475.7  & 1.79   &178.2 \\
      9    &   Gwangpan        &	358 	 &  18    &       186     & 353.9  & 4.05   &177.6 \\
      10   &   Gunja 1         &	474 	 &  51    &       220     & 467.0  & 2.34   &179.5 \\
      11   &   Gunja 2         &	885 	 &  120   &       255     & 879.8  & 1.88   &180.1 \\
      12   &   Dongsan 1       &	680	 &  86    &       295     & 684.8  & 1.73   &180.8 \\
      13   &   Dongsan 2       &	1113	 &  120.5 &       273     & 1146.7 & 1.42   &179.6 \\
      14   &   Bukbang 1       &	2307	 &  152   &       252     & 2339.7 & 1.38   &180.0 \\
      15   &   Bukbang 2       &	326	 &  65    &       235     & 321.5  & 1.64   &179.7 \\
      16   &   Bukbang 3       &	1518	 &  159   &       246     & 1533.7 & 1.34   &172.1 \\
      17   &   Hwachon 1       &	721	 &  85    &       216     & 716.7  & 1.51   &176.7 \\
      18   &   Hwachon 2       &	378	 &  80    &       200     & 378.5  & 1.26   &177.4 \\
      19   &   Hwachon 3       &	527	 &  61    &       204     & 517.9  & 2.10   &178.5 \\
      20   &   Hwachon 4       &	35	 &  3	  &       252     & 36.8   & 21.10  &182.2 \\
      21   &   Hwachon 5       &	125	 &  10    &       264     & 118.4  & 2.34   &182.3 \\
      22   &   Hwachon 6       &	629	 &  41	  &       283     & 594.4  & 2.25   &184.7 \\
      23   &   Hwachon 7       &	40	 &  3	  &       308     & 37.2   & 18.50  &187.6 \\
      24   &   Hwachon 8       &	978	 &  124   &       328     & 964.1  & 1.82   &182.9 \\
      25   &   Hwachon 9       &	3705	 &  304   &       342     & 3725.1 & 1.18   &189.8 \\
      26   &   Naechon 1       &	1261	 &  52.5  &       332     & 1193.6 & 1.82   &180.2 \\
      27   &   Naechon 2       &	88	 &  4     &       309     & 84.8   & 10.30  &182.9 \\
      28   &   Naechon 3       &	135	 &  8.5   &       354     & 146.9  & 4.04   &182.3 \\
      29   &   Naechon 4       &	355	 &  39    &       362     & 356.6  & 3.10   &184.6 \\
      30   &   Naechon 5       &	205	 &  19.5  &       393     & 179.3  & 3.90   &188.8 \\
      31   &   Seoseok         &	3061	 &  221.5 &       430     & 3122.0 & 1.24   &181.0 \\
      32   &   Haengchiryeong  &	1422	 &  104   &       497     & 1424.3 & 1.23   &191.3 \\
      33   &   Sangnam 1       &	60	 &  3     &       520     & 59.9   & 28.70  &183.1 \\
      34   &   Sangnam 2       &	115	 &  5     &       527     & 102.3  & 21.10  &186.8 \\
      35   &   Sangnam 3       &	1719	 &  126.5 &       485     & 1710.6 & 1.25   &191.7 \\
      36   &   Sangnam 4       &	1434	 &  195   &       454     & 1444.6 & 1.20   &192.7 \\
      37   &   Sangnam 5       &	1773	 &  149   &       442     & 1762.8 & 1.25   &198.4 \\
      38   &   Sangnam 6       &	1816	 &  100   &       405     & 1786.4 & 1.59   &188.3 \\
      39   &   Sangnam 7       &	2278	 &  264   &       381     & 2338.8 & 1.22   &187.4 \\
      40   &   Girin 1         &	114	 &  7	  &       408     & 97.3   & 2.60   &179.5 \\
      41   &   Girin 2         &	397	 &  41    &       398     & 375.8  & 2.27   &179.9 \\
      42   &   Girin 3         &	162	 &  11.5  &       402     & 157.9  & 2.00   &179.6 \\
      43   &   Girin 4         &	750	 &  128.5 &       418     & 762.3  & 1.25   &183.9 \\
      44   &   Girin 5         &	1148	 &  109.5 &       441     & 1169.2 & 1.48   &190.7 \\
      45   &   Girin 6         &	2665	 &  288.5 &       432     & 2757.8 & 1.15   &190.0 \\
      46   &   Inje-Yangyang   &	10962	 &  440   &       335     & 11543.3& 1.06   &176.1 \\
      47   &   Seomyeon 2      &	248	 &  13.5  &       217     & 236.0  & 4.67   &175.8 \\
      \hline  
  \end{tabular}
  \end{center}
\end{table*}

\end{document}